\documentclass[12pt]{article}
\usepackage{amsfonts}
\usepackage{amsmath}
\usepackage{amssymb}
\usepackage{mathptmx}
\usepackage{geometry}
\usepackage{graphicx}
\usepackage{hyperref}
\usepackage{cancel}
\usepackage{textcomp}
\usepackage{tikz}
\usepackage{bm}
  \usepackage{mathrsfs}
  \usepackage{filecontents}
\usepackage{times}
\usepackage{epsfig}
\usepackage{xcolor}
\usepackage{slashed}
\usepackage{booktabs}% build a table
\usepackage{latexsym}
\usepackage{verbatim}
\usepackage{extarrows}
\usepackage{multirow}
\usepackage{rotating}
\usepackage{colortbl}
\usepackage{indentfirst}
\usepackage{float}
\definecolor{mygray}{gray}{.9}
\definecolor{intnull}{RGB}{213,229,255}
\usepackage{arydshln}
\usepackage{diagbox}
\usepackage{makecell}
\usepackage{array}
\usepackage{cite}
\usepackage{caption}
\usepackage{tikz}
\usetikzlibrary{arrows,shapes,chains}
\usepackage{graphicx, subfig}
\begin{document}
\renewcommand{\thefootnote}{\fnsymbol{footnote}}
\baselineskip=16pt
\pagenumbering{arabic}
\vspace{1.0cm}
\begin{center}
{\Large\sf  Quasinormal modes and shadows of a new family of Ay\'{o}n-Beato-Garc\'{\i}a black holes}
\\[10pt]
\vspace{.5 cm}

{Xin-Chang Cai\footnote{E-mail address: caixc@mail.nankai.edu.cn} and Yan-Gang Miao\footnote{Corresponding author. E-mail address: miaoyg@nankai.edu.cn}}
\vspace{3mm}

{School of Physics, Nankai University, Tianjin 300071, China}

\vspace{4.0ex}

{\bf Abstract}
\end{center}

We obtain a type of Ay\'{o}n-Beato-Garc\'{\i}a (ABG) related black hole solutions with five parameters: the mass $m$, the charge $q$, and three dimensionless parameters $\alpha$, $\beta$ and $\gamma$ associated with nonlinear electrodynamics. We find that this type of black holes is regular under the conditions: $\alpha \gamma \geqslant 6$, $\beta \gamma \geqslant 8$, and $\gamma >0$. Here we focus on the saturated case: $\alpha={6}/{\gamma}$ and $\beta ={8}/{\gamma }$, such that only three parameters $m$, $q$ and $\gamma$ remain, which leads to a new family of ABG black holes. For such a family of black holes, we investigate the influence of the charge $q$ and the parameter $\gamma$ on the horizon radius and the Hawking temperature. In addition, we calculate the quasinormal mode frequencies of massless scalar field perturbations by using the sixth-order WKB approximation method and the unstable circular null geodesic method in the eikonal limit. On the one hand, our results show that the increase of the charge $q$ makes the scalar waves decay faster at first and then slowly except for the case of $\gamma=2$. On the other hand, they show that the increase of the parameter $\gamma$ makes the scalar waves decay at first sharply and then slowly. In particular, $\gamma=1$ can be regarded as the critical value for the transition from an unstable configuration to a stable one. Finally, we compute the shadow radius for the new family of ABG black holes  and  use the shadow data  of  the $M87^{*}$ black  hole detected by the Event Horizon Telescope to  provide  an upper limit on the charge $q$  of  the  new black holes. We find that the increase of the charge $q$  makes the shadow radius decrease monotonically, while the increase of the parameter $\gamma$ makes the shadow radius increase at first rapidly and then almost remain unchanged, especially the parameter $\gamma$ has a significant impact on the shadow radius when it is less than one.  Moreover, using the shadow data  of the $M87^{*}$ black  hole, we find that the upper limit of  the charge $q$ increases rapidly at first and then slowly but does not exceed the mass of the  $M87^{*}$  black hole at last when the parameter $\gamma$  is increasing  and going to infinity, and that  the data restrict the frequency range of the fundamental mode with  $l=1$ to $1.4\times 10^{-6}Hz\sim 1.9\times 10^{-6} Hz$.

\newpage

\section{Introduction}

It is well known that the black hole solutions of Einstein's field equations have a spacetime singularity which cannot be eliminated by coordinate transformations. In fact, the singularity theorem proved by Penrose and Hawking~\cite{P1,P2} shows that the existence of singularities in general relativity is inevitable under some reasonable physical conditions. Since the spacetime singularity inside a black hole can lead to the failure of all physical laws, Penrose proposed~\cite{P3,P4} the famous cosmic supervision hypothesis in 1969: The spacetime singularity is always hidden by the event horizon of black holes. However, it is widely believed~\cite{P5} that spacetime singularities are unphysical objects that should not exist in nature due to limitations of classical theories of gravity, and that these singularities can be avoided especially when quantum effects are taken into account. Based on this idea, Bardeen proposed~\cite{P6} the first static spherically symmetric regular black hole solution with an unknown physical source. Later, Ay\'{o}n-Beato and Garc\'{\i}a realized~\cite{P7} that the physical source of regular black holes may be a nonlinear electromagnetic field, and under the framework of general relativity, they successfully obtained~\cite{P7,P8,P9,P10} a series of regular black hole solutions by coupling an appropriate field of nonlinear electrodynamics to Einstein's field equations. In this way, many regular black hole solutions have been obtained~\cite{P5,P11,P12,P13} so far.

Inspired by Ref.~\cite{P5}, we notice that the metric function $f(r)$ of regular black hole solutions with magnetic charges or electric charges can contain a term like ${r^{a}}/{(r^{b}+q^{b})^{{a}/{b}}}$, where $a$ and $b$ are two dimensionless parameters associated with nonlinear electrodynamics, and then obtain a type of ABG related black hole solutions. Within this framework of black hole solutions, the ABG black hole solution~\cite{P7} and its generalization~\cite{P10} become the special cases of this type of ABG related black hole solutions. %in which the term $\frac{r^{a}}{(r^{b}+q^{b})^{{a}/{b}}}$ appears in their metric functions.
Our fulfilment may be a further improvement of the ABG black hole and its generalized solutions.
%In other words, whether we can construct a more general ABG black hole solution , or not. The answer is yes. In addition, the more general ABG black hole solution should also be a further improvement of the ABG related solutions.

Recently, the gravitational waves from the merger of two black holes detected by LIGO and Virgo collaborations~\cite{P14,P15} and the first black hole image detected by the Event Horizon Telescope (EHT) collaboration~\cite{P16,P17} have greatly stimulated the enthusiasm for black hole physics. The two achievements represent the significance of black hole perturbations and black hole shadows, respectively. So far, the quasinormal modes and the shadows of a large number of black hole models have been widely and deeply analyzed, for instance, see Refs.~\cite{P18,P19,P20,P21,P22,P23,P24,P25,P26,P27,P28,P29,P30,P31,P32,P33,P34,P35,P36,P37,P38,P61,P62,P63,P39,P40,
P41,P42,P43,P44,P45,P46,P47,P48}. We hope the construction of a type of ABG related black hole solutions will expand the family of ABG black holes and promote the related studies in quasinormal modes and shadows.

The paper is organized as follows. In Sec. 2, we construct a type of ABG related black hole solutions. Then, we investigate in Sec. 3 the characteristics of its saturated form --- a new family of ABG black holes. We compute the quasinormal mode frequencies of massless scalar field perturbations in Sec. 4 and the shadows in Sec. 5 for this new family of black holes.  Finally, we make a simple summary in Sec. 6. We use the units $c=G=k_{B}=\hbar=1$ and the sign convention $(-,+,+,+)$ throughout this paper.

\section{Construction of a type of ABG related black holes}

The action of Einstein's gravity coupled to a nonlinear electrodynamic field takes~\cite{P7,P8,P9,P10} the form,
\begin{equation}
\label{1}
S=\int d^{4}x \sqrt{-g}\left[\frac{1}{16\pi }R-\frac{1}{4\pi }L(P) \right ].
\end{equation}
Here $R$ is the scalar curvature, $L(P)=2PH_{P}-H(P)$ is a function of the invariant $P\equiv \frac{1}{4}P_{\mu \nu }P^{\mu \nu }$, and the antisymmetric tensor $P_{\mu \nu}\equiv {F_{\mu \nu}}/{H_{P}}$,
where $H_{P}\equiv {\mathrm{d} H(P)}/{\mathrm{d} P}$, $H(P)$ is the structure function of the nonlinear electrodynamic theory corresponding to the nonlinear electromagnetic tensor $F_{\mu \nu }\equiv \partial _{\mu }A_{\nu }- \partial _{\nu }A_{\mu }$ with the electromagnetic potential $A_{\mu }$.

Varying Eq.~(\ref{1}) with respect to the metric $g_{\mu \nu }$ and the electromagnetic potential $A_{\mu }$, respectively, one can derive~\cite{P7,P8,P9,P10} the Einstein field equations coupled with nonlinear electrodynamics as follows,
\begin{equation}
\label{2}
G_{\mu }^{\nu }=2[H_{P}P_{\mu \lambda }P^{\nu \lambda }-\delta _{\mu }^{\nu }(2PH_{P}-H(P))],
\end{equation}
\begin{equation}
\label{3}
\bigtriangledown _{\mu }P^{\nu \mu }=0.
\end{equation}
A static spherically symmetric black hole can be described by the line element,
\begin{equation}
\label{4}
ds^{2}=-f(r)dt^{2}+\frac{dr^{2}}{f(r)}+r^{2}\left(d\theta^{2}+\sin^{2}\theta d\varphi^{2}\right),
\end{equation}
where $f(r)$ is the metric function. By assuming $P_{\mu \nu }=2\delta _{[\mu }^{t}\delta _{\nu]}^{r}D(r)$, where $D(r)$ is a function of $r$,  and combining the line element Eq.~(\ref{4}) with Eq.~(\ref{3}), one can get\cite{P7,P8,P10}
\begin{equation}
\label{5}
P_{\mu \nu }=2\delta _{[\mu }^{t}\delta _{\nu]}^{r}\frac{q}{r^{2}} \qquad \Longrightarrow  \qquad    P=-\frac{D^{2}}{2}=-\frac{q^{2}}{2r^{4}},
\end{equation}
where $q$ is integral constant that acts as electric charge. Note that $q$ specifically refers to as $\left | q \right |$ since parameter $\gamma $ appeared in the formula below may take an odd number.
We choose such a structure function,
\begin{equation}
\begin{aligned}
\label{6}
H(P)=\frac{P \left[1-\left(\frac{\beta  \gamma }{2}-1\right) \left(-2 P q^2\right)^{\gamma /4}\right]}{\left[\left(-2 P q^2\right)^{\gamma /4}+1\right]^{\frac{\beta}{2}+1}}-\frac{\alpha  \gamma  m \left(-2 P q^2\right)^{\frac{\gamma +3}{4}}}{2 q^3 \left[\left(-2 P q^2\right)^{\gamma /4}+1\right]^{\frac{\alpha}{2}+1}},
\end{aligned}
\end{equation}
where $\alpha$, $\beta$, and $\gamma$ are three dimensionless parameters and $m$ is a parameter related to mass. If  $\gamma>1$, for the weak field limit, $P\ll 1$, we have $H\approx P$, which means that the nonlinear electrodynamics returns to Maxwell's theory; if $0<\gamma<1$, we have $H\approx -{\alpha  \gamma  m \left(-2 P q^2\right)^{{(\gamma +3)}/{4}}}/{(2 q^3)}$ under the weak field limit, which implies that the nonlinear electrodynamics does not return to Maxwell's theory. Therefore,  $\gamma=1$ is an important critical value of this nonlinear theory.

By substituting Eqs.~(\ref{4})-(\ref{6}) into Eq.~(\ref{2}), we obtain the metric function,
\begin{equation}
\label{7}
f(r)=1-\frac{2 m r^{\frac{\alpha  \gamma }{2}-1}}{\left(q^{\gamma }+r^{\gamma }\right)^{\alpha /2}}+\frac{q^2 r^{\frac{\beta  \gamma }{2}-2}}{\left(q^{\gamma }+r^{\gamma }\right)^{\beta /2}},
\end{equation}
which is a type of Ay\'{o}n-Beato-Garc\'{\i}a (ABG) related black hole solutions. It is easy to see that this type of solutions returns to the ABG black hole solution~\cite{P7}when $\alpha=3$, $\beta=4$, and $\gamma=2$, and to the generalized ABG black hole solution~\cite{P10} when $\gamma=2$.

Combining Eq.~(\ref{4}) and Eq.~(\ref{7}), we can calculate the Ricci scalar of this type of ABG related black holes,
\begin{equation}
\label{8}
R=\frac{\alpha  \gamma  m q^{\gamma} r^{\frac{\alpha  \gamma }{2}-3} \left[(\alpha  \gamma +2) q^{\gamma}-2 (\gamma -1) r^{\gamma}\right]}{2 \left(r^{\gamma }+q^{\gamma }\right)^{\frac{\alpha }{2}+2}}-\frac{\beta  \gamma  q^{\gamma +2} r^{\frac{\beta  \gamma}{2}-4} \left[(\beta  \gamma -2) q^{\gamma}-2 (\gamma +1) r^{\gamma}\right]}{4 \left(r^{\gamma}+q^{\gamma}\right)^{\frac{\beta }{2}+2}}.
\end{equation}
Similarly, we can also obtain the curvature invariants, $R_{\mu \nu }R^{\mu \nu }$  and  $R_{\mu \nu \rho \sigma  }R^{\mu \nu \rho \sigma}$. We conclude that the three invariants converge at $r=0$ under the conditions: $\alpha \gamma \geqslant 6$, $\beta \gamma \geqslant 8$, and $\gamma >0$, which means that the type of ABG related black holes is regular.

If the above conditions are saturated, i.e. $\alpha=\frac{6}{\gamma}$ and  $\beta =\frac{8}{\gamma}$, we give a new family of ABG black holes with the metric function,
\begin{equation}
\label{9}
f(r)=1-\frac{2 m r^2}{\left(r^{\gamma }+q^{\gamma }\right)^{3/\gamma }}+\frac{q^2 r^2}{\left(r^{\gamma }+q^{\gamma }\right)^{4/\gamma }},
\end{equation}
which contains three parameters, $m$, $q$, and $\gamma$, and satisfies the weak energy condition, i.e. $H(P)<0$ and  $H_{P}>0$. In addition, Eq.~(\ref{9}) returns to the metric function of the Reissner-Nordstr\"om black hole solution when $\gamma \to \infty $.

\section{Characteristics of the new family of ABG black holes}

By using Eq.~(\ref{9}) and the formula of Hawking temperature~\cite{P49,P50}, $T_{\rm H}=\frac{1}{4\pi}f^{\prime}(r_{\rm H})$,  we can calculate the Hawking temperature for the new family of ABG black holes,
\begin{equation}
\label{10}
T_{\rm H}=\frac{r_{\rm H}^{\gamma} \left[1-q^2 r_{\rm H}^2 \left(r_{\rm H}^{\gamma}+q^{\gamma}\right)^{-4/\gamma }\right]-2 q^{\gamma}}{4 \pi  r_{\rm H} \left(r_{\rm H}^{\gamma}+q^{\gamma}\right)},
\end{equation}
where $r_{\rm H}$ is the event horizon radius.

In Fig. 1, we plot the metric function $f(r)$, Eq.~(\ref{9}), with respect to $r$ for different values of charge $q$ when $\gamma$ is given for different cases. We can see that this family of black holes may have none, or one, or two horizons as the charge $q$ decreases, which is just like the behaviour of the ABG black hole and  Reissner-Nordstr\"om black hole. We note that the smaller the parameter $\gamma$ is, the smaller the charge $q$ of the extreme black hole is, see the values of $q$ corresponding to blue curves.
%, for instance, $q_{max}\sim 10^{-4}$ if $\gamma=0.2$. Therefore, we only calculate the case where $\gamma \geqslant 0.5$.

\begin{figure}
		\centering
		\begin{minipage}{.5\textwidth}
			\centering
			\includegraphics[width=80mm]{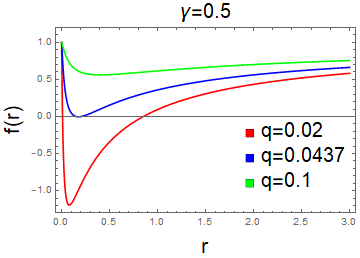}
		\end{minipage}%
		\begin{minipage}{.5\textwidth}
			\centering
			\includegraphics[width=80mm]{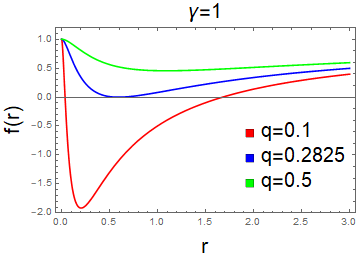}
		\end{minipage}
\begin{minipage}{.5\textwidth}
			\centering
			\includegraphics[width=80mm]{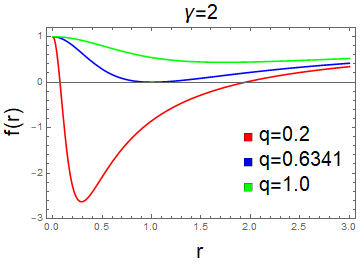}
		\end{minipage}%
		\begin{minipage}{.5\textwidth}
			\centering
			\includegraphics[width=80mm]{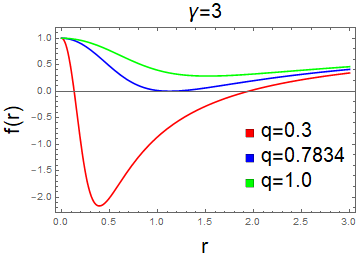}
		\end{minipage}
\begin{minipage}{.5\textwidth}
			\centering
			\includegraphics[width=80mm]{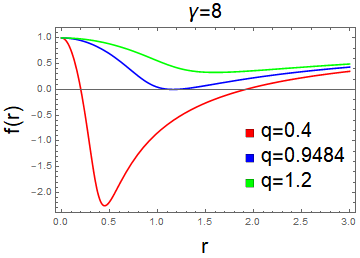}
		\end{minipage}%
		\begin{minipage}{.5\textwidth}
			\centering
			\includegraphics[width=80mm]{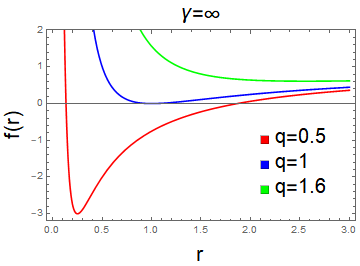}
		\end{minipage}
		\caption*{Fig. 1. Function $f(r)$ with respect to $r$ for different values of charge $q$ when parameter $\gamma$ is given for different cases, where the case of $\gamma=2$ corresponding to the ABG black hole and
the case of $\gamma \to \infty$  to the Reissner-Nordstr\"om black hole are attached for comparison. Here we set $m=1$.}
\label{figure1}
\end{figure}

In Fig. 2, we plot the extreme event horizon radius  $r_{\rm EH}$ with respect to the parameter $\gamma$. We can see that the increase of the parameter $\gamma$
makes the extreme event horizon radius increase at first and then decrease, especially there exists a special value of parameter $\gamma$ that maximizes the extreme event horizon radius, which is associated with the structure of the nonlinear electrodynamics, see the analysis of the structure function in section 2.

\begin{figure}
\centering
\begin{minipage}[t]{0.8\linewidth}
\centering
\includegraphics[width=100mm]{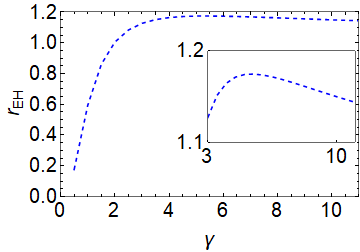}
\caption*{Fig. 2. The extreme event horizon radius $r_{\rm EH}$ with respect to the parameter $\gamma$. Here we set $m=1$.}
\label{fig2}
\end{minipage}
\end{figure}

In Fig. 3, we plot the Hawking temperature $T_{\rm H}$ with respect to the event horizon radius $r_{\rm H}$ for different values of charge $q$ when $\gamma$ is given for different cases.
We can see that the increase of the event horizon radius $r_{\rm H}$ makes the Hawking temperature $T_{\rm H}$ increase at first and then decrease, and the increase of the charge $q$ makes the Hawking temperature decrease.  This behaviour is just like that of the ABG black hole and Reissner-Nordstr\"om black hole. In addition, we can also see that the Hawking temperature increases more and more slowly with the increasing of parameter $\gamma$ for a fixed $q$.

\begin{figure}
		\centering
		\begin{minipage}{.5\textwidth}
			\centering
			\includegraphics[width=80mm]{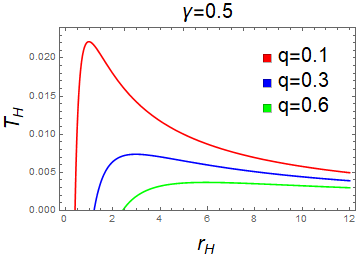}
		\end{minipage}%
		\begin{minipage}{.5\textwidth}
			\centering
			\includegraphics[width=80mm]{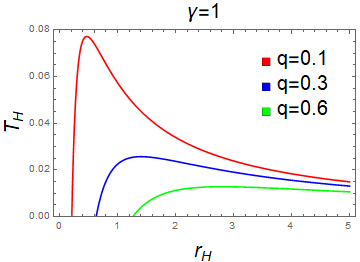}
		\end{minipage}
\begin{minipage}{.5\textwidth}
			\centering
			\includegraphics[width=80mm]{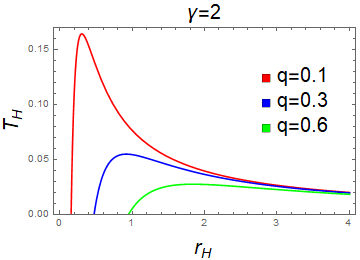}
		\end{minipage}%
		\begin{minipage}{.5\textwidth}
			\centering
			\includegraphics[width=80mm]{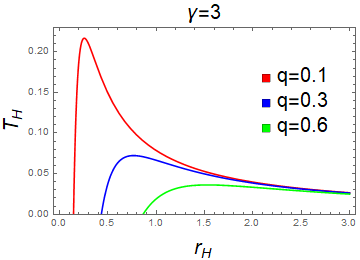}
		\end{minipage}
\begin{minipage}{.5\textwidth}
			\centering
			\includegraphics[width=80mm]{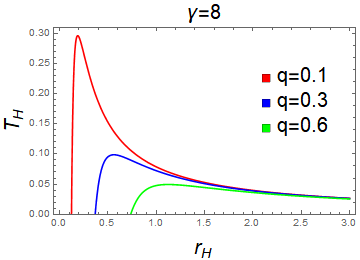}
		\end{minipage}%
		\begin{minipage}{.5\textwidth}
			\centering
			\includegraphics[width=80mm]{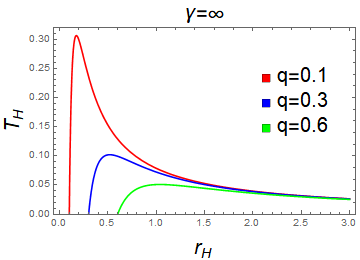}
		\end{minipage}

		\caption*{Fig. 3. The Hawking temperature $T_{\rm H}$ with respect to the event horizon radius $r_{\rm H}$ for different values of charge $q$ when parameters $\gamma$ is given for different cases, where the case of $\gamma=2$ corresponding to the ABG black hole and the case of $\gamma\to \infty$ to the Reissner-Nordstr\"om black hole are attached for comparison.}
\label{figure3}
\end{figure}

\section{Quasinormal mode frequencies of massless scalar field perturbations for the new family of ABG black holes}

In this section, we study the quasinormal modes of a neutral massless scalar field perturbation around one of the  new ABG black holes. The propagation of the massless scalar field $\Phi $ in the curved spacetime depicted by Eq.~(\ref{9}) is described by the Klein-Gordon equation,
\begin{equation}
\label{11}
\frac{1}{\sqrt{-g}}\partial _{\mu }(\sqrt{-g}g^{\mu \nu }\partial _{\nu }\Phi )=0,
\end{equation}
where $g$ is the determinant of  the metric tensor $g_{\mu \nu }$. By defining the tortoise coordinate $dr_{*}\equiv \frac{dr}{f(r)}$ and substituting $\Phi =e^{-i\omega t}Y_{lm}(\theta,\varphi)\frac{\Psi (r)}{r}$ together with  Eqs.~(\ref{4}) and (\ref{9}) into Eq.~(\ref{11}), we obtain
\begin{equation}
\label{12}
\frac{\mathrm{d} ^{2}\Psi (r_{*})}{\mathrm{d} r^{2}_{*}}+[\omega ^{2}-V(r)]\Psi (r_{*})=0,
\end{equation}
with the effective potential,
\begin{equation}
\label{13}
V(r)=\left[1-\frac{2 m r^2}{\left(q^{\gamma }+r^{\gamma }\right)^{3/\gamma }}+\frac{q^2 r^2}{\left(q^{\gamma }+r^{\gamma }\right)^{4/\gamma }}\right] \left\{\frac{(l+1) l}{r^2}+\frac{2 \left[m \left(r^{\gamma }-2 q^{\gamma }\right) \left(q^{\gamma }+r^{\gamma }\right)^{1/\gamma }+q^2 \left(q^{\gamma }-r^{\gamma }\right)\right]}{\left(q^{\gamma }+r^{\gamma }\right)^{\frac{\gamma +4}{\gamma }}}\right\},
\end{equation}
where $\omega$ is complex quasinormal mode frequency and $l$ multipole number.

\subsection{Quasinormal mode frequencies of massless scalar field perturbations by the sixth-order WKB  approximation}

We use the WKB approximation method to calculate numerically the quasinormal mode frequencies of massless scalar field perturbations for the new family of ABG black holes. This method has been developed~\cite{P52,P53,P54,P55} from the first order to the latest 13th order. We choose the sixth-order WKB approximation method for the sake of the computational efficiency and accuracy. Then, we draw the graphs of real parts and negative imaginary parts of quasinormal mode frequencies with respect to the charge $q$ and the parameter $\gamma$, as shown in Fig. 4  and  Fig. 5, respectively. We note that the charge $q$ in Fig. 4 ends with the values corresponding to the extreme configurations as shown by the blue curves in Fig. 1.

\begin{figure}
		\centering
		\begin{minipage}{.5\textwidth}
			\centering
			\includegraphics[width=80mm]{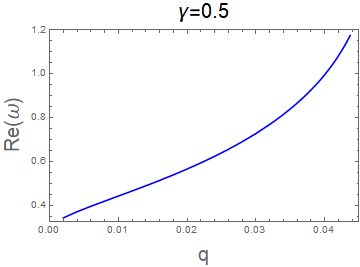}
		\end{minipage}%
		\begin{minipage}{.5\textwidth}
			\centering
			\includegraphics[width=80mm]{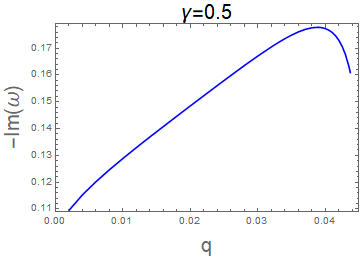}
		\end{minipage}
\begin{minipage}{.5\textwidth}
			\centering
			\includegraphics[width=80mm]{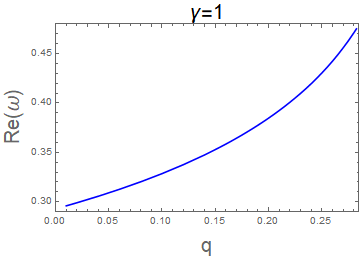}
		\end{minipage}%
		\begin{minipage}{.5\textwidth}
			\centering
			\includegraphics[width=80mm]{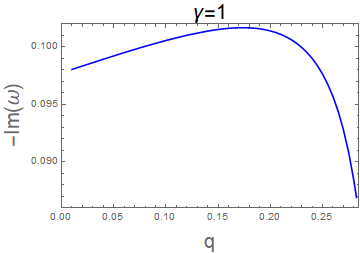}
		\end{minipage}
\begin{minipage}{.5\textwidth}
			\centering
			\includegraphics[width=80mm]{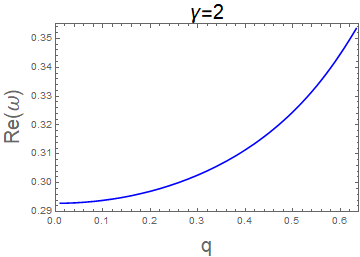}
		\end{minipage}%
		\begin{minipage}{.5\textwidth}
			\centering
			\includegraphics[width=80mm]{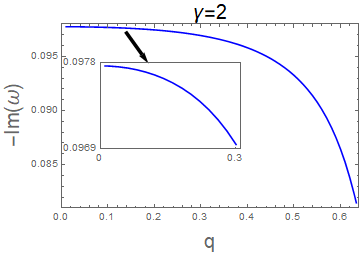}
		\end{minipage}
\end{figure}
\begin{figure}
		\centering
\begin{minipage}{.5\textwidth}
			\centering
			\includegraphics[width=80mm]{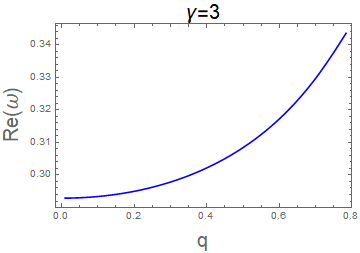}
		\end{minipage}%
		\begin{minipage}{.5\textwidth}
			\centering
			\includegraphics[width=80mm]{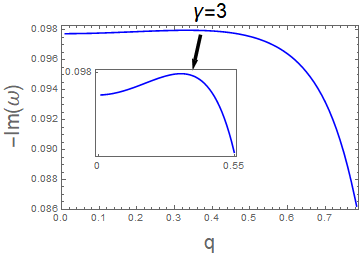}
		\end{minipage}
\begin{minipage}{.5\textwidth}
			\centering
			\includegraphics[width=80mm]{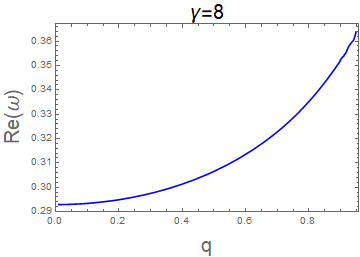}
		\end{minipage}%
		\begin{minipage}{.5\textwidth}
			\centering
			\includegraphics[width=80mm]{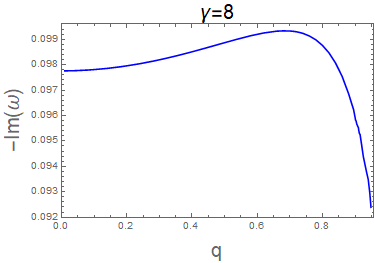}
		\end{minipage}
\begin{minipage}{.5\textwidth}
			\centering
			\includegraphics[width=80mm]{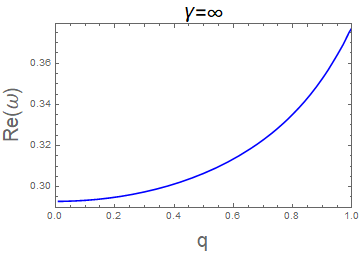}
		\end{minipage}%
		\begin{minipage}{.5\textwidth}
			\centering
			\includegraphics[width=80mm]{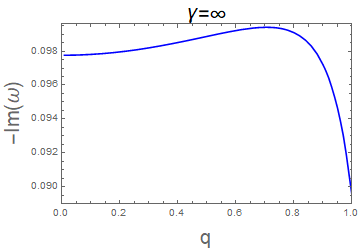}
		\end{minipage}
\caption*{Fig. 4. The real parts and negative imaginary parts of quasinormal mode frequencies of massless scalar  field perturbations with respect to the charge $q$ when $\gamma$ is given for different cases, where the case of $\gamma=2$ corresponding to the ABG black hole and the case of $\gamma\to\infty$ to the Reissner-Nordstr\"om black hole are attached for comparison. Here we set $m=1$, $l=1$, and $n=0$.}
\label{figure4}
\end{figure}

From Fig. 4 we can see that the behaviour of quasinormal modes of the new family of ABG black holes is similar to  that of the Reissner-Nordstr\"om black hole as the charge $q$ increases except for the case of $\gamma=2$ --- the ABG black hole. That is, the real parts increase monotonically, while the negative imaginary parts increase at first to the maximum and then decrease for the new family of ABG black holes, which means that the scalar wave decays fast at the beginning and then slowly as the charge $q$ increases except for the ABG black hole.% This also indicates that the ABG black hole is a more special regular black hole among the special  generalized ABG black holes.

\begin{figure}
		\centering
\begin{minipage}{.5\textwidth}
			\centering
			\includegraphics[width=80mm]{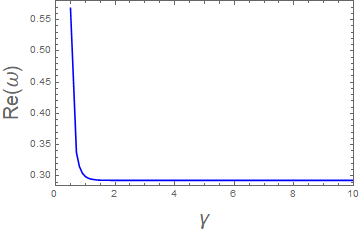}
		\end{minipage}%
		\begin{minipage}{.5\textwidth}
			\centering
			\includegraphics[width=80mm]{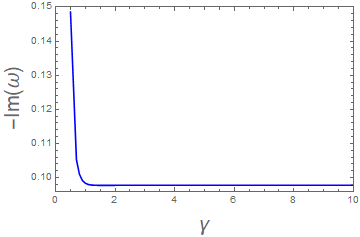}
		\end{minipage}
\caption*{Fig. 5. The real parts and negative imaginary parts of quasinormal mode frequencies of massless scalar  field perturbations with respect to the parameters $\gamma$. Here we set $m=1$, $q=0.02$, $l=1$,  and $n=0$.}
\label{figure5}
\end{figure}

From Fig. 5 we can see that both the real parts and negative imaginary parts decrease sharply and then almost remain unchanged as the parameter $\gamma$ increases, which shows that the scalar wave decays quickly at first and then slowly. This phenomenon implies that the parameter $\gamma$  has a significant impact on the quasinormal mode frequencies when $0<\gamma<1$. In particular, $\gamma=1$ can be regarded as the critical value for the transition from an unstable to a stable configuration of new ABG black holes because $\gamma=1$ is an important critical value of the nonlinear electrodynamics as mentioned in section 2.

\subsection{Quasinormal mode frequencies in the eikonal limit calculated via the unstable circular null geodesics}

The unstable circular null geodesic method was first proposed by Cardoso et al.~\cite{P56}for calculation of the quasinormal mode frequencies of a static spherically symmetric black hole in the eikonal limit, $ l\gg 1$. The effective potential Eq.~(\ref{13}) reduces to the following form in this limit,
\begin{equation}
\label{16}
V(r)= \left[1-\frac{2 m r^2}{\left(q^{\gamma }+r^{\gamma }\right)^{3/\gamma }}+\frac{q^2 r^2}{\left(q^{\gamma }+r^{\gamma }\right)^{4/\gamma }}\right] \frac{l^{2}}{r^{2}},
\end{equation}
and the corresponding quasinormal mode frequencies are determined~\cite{P56,P57,P58} by
\begin{equation}
\label{17}
\omega _{l\gg 1}=l\Omega_c -i\left(n+\frac{1}{2}\right)|\lambda_{\rm L}|,
\end{equation}
with
\begin{equation}
\label{18}
\Omega _c=\frac{\sqrt{f(r_{c})}}{r_{c}},        \qquad   \lambda_{\rm L} =\sqrt{\frac{f(r_{c})\left[2f(r_{c})-r_{c}^{2}{f}''(r_{c})\right]}{2r_{c}^{2}}},
\end{equation}
where the angular velocity $\Omega_c $ and the Lyapunov exponent $\lambda_{\rm L}$ control the real parts and imaginary parts of the quasinormal mode frequencies $\omega$, respectively.  Moreover, the radius $r_{c}$ of circular null geodesics is given by
\begin{equation}
\label{19}
2f(r_{c})-r_{c}\left.\frac{\mathrm{d} f(r)}{\mathrm{d} r}\right|_{r=r_{c}}=0.
\end{equation}

As discussed in the above subsection, we plot the graphs of the angular velocity $\Omega_c$ and the Lyapunov exponent $\lambda_{\rm L}$ with respect to the charge $q$ and the parameter $\gamma$ in Fig. 6 and  Fig. 7, respectively. By comparing Fig. 4 with Fig. 6 and Fig. 5 with Fig. 7, we conclude that the two methods coincide with each other.

%From Fig. 6  we can see that  except for the  ABG black hole, this special  generalized ABG black hole is similar to the  Reissner-Nordstr\"om black hole in that as the increase of  the charge $q$, the angular velocity $\Omega_c $ monotonically increase, while the Lyapunov exponent $\lambda_{\rm L}$  first increase to the maximum and then decrease, which  means that as the increase of  the charge $q$, real parts  of  quasinormal frequencies   at the eikonal  limit  monotonically increase, while negative imaginary parts first increase to the maximum and then decrease.

%From Fig. 7  we can see that  as the  parameter $\gamma$  increases, both the angular velocity $\Omega_c $ and the Lyapunov exponent $\lambda_{\rm L}$   decrease sharply and then almost remains unchanged, which means that as  the  parameter $\gamma$   increases, real parts  and negative imaginary parts  of  quasinormal frequencies  at the eikonal  limit  decrease sharply and then almost remains unchanged. In particular, in the parameter range of less than 1,  the  parameter $\gamma$  has a particularly significant impact on the quasinormal frequencies   at this  limit.

\begin{figure}
		\centering
		\begin{minipage}{.5\textwidth}
			\centering
			\includegraphics[width=80mm]{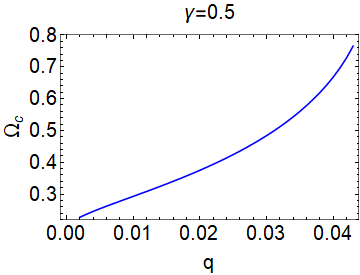}
		\end{minipage}%
		\begin{minipage}{.5\textwidth}
			\centering
			\includegraphics[width=80mm]{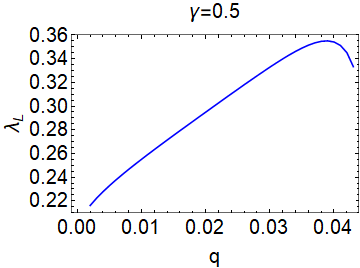}
		\end{minipage}
\begin{minipage}{.5\textwidth}
			\centering
			\includegraphics[width=80mm]{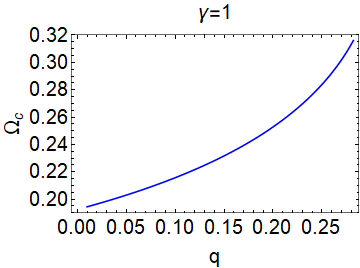}
		\end{minipage}%
		\begin{minipage}{.5\textwidth}
			\centering
			\includegraphics[width=80mm]{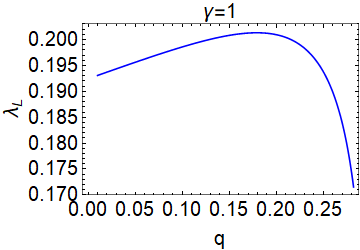}
		\end{minipage}
\begin{minipage}{.5\textwidth}
			\centering
			\includegraphics[width=80mm]{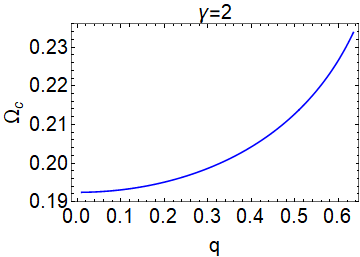}
		\end{minipage}%
		\begin{minipage}{.5\textwidth}
			\centering
			\includegraphics[width=80mm]{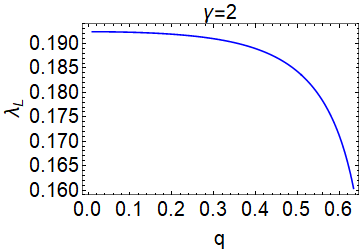}
		\end{minipage}
\end{figure}

\begin{figure}
		\centering
\begin{minipage}{.5\textwidth}
			\centering
			\includegraphics[width=80mm]{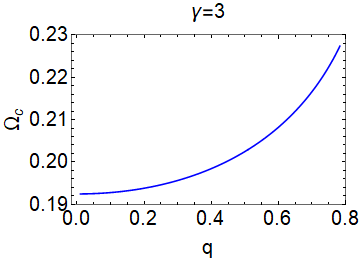}
		\end{minipage}%
		\begin{minipage}{.5\textwidth}
			\centering
			\includegraphics[width=80mm]{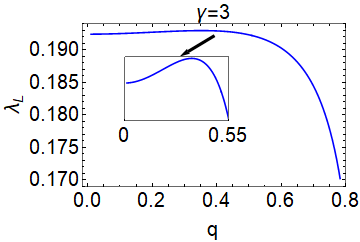}
		\end{minipage}
\begin{minipage}{.5\textwidth}
			\centering
			\includegraphics[width=80mm]{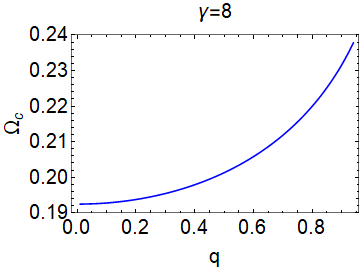}
		\end{minipage}%
		\begin{minipage}{.5\textwidth}
			\centering
			\includegraphics[width=80mm]{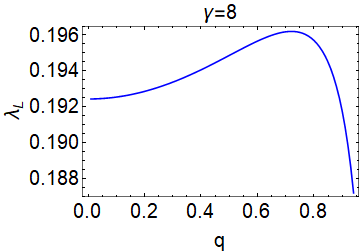}
		\end{minipage}
\begin{minipage}{.5\textwidth}
			\centering
			\includegraphics[width=80mm]{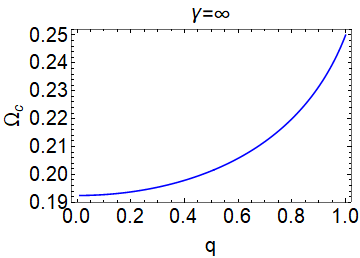}
		\end{minipage}%
		\begin{minipage}{.5\textwidth}
			\centering
			\includegraphics[width=80mm]{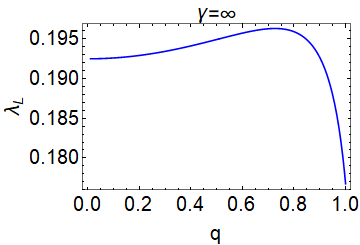}
		\end{minipage}

		\caption*{Fig. 6.  The angular velocity $\Omega_c $ and the Lyapunov exponent $\lambda_{\rm L}$ with respect to the charge $q$ when $\gamma$ is given for different cases, where the case of $\gamma=2$ corresponding to the ABG black hole and the case of $\gamma\to\infty$ to the Reissner-Nordstr\"om black hole are attached for comparison. Here we set $m=1$.}
\label{figure4}
\end{figure}

\begin{figure}
		\centering
\begin{minipage}{.5\textwidth}
			\centering
			\includegraphics[width=80mm]{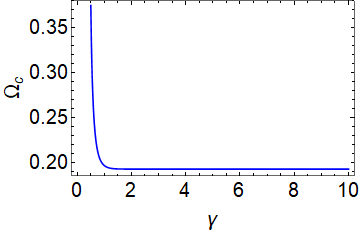}
		\end{minipage}%
		\begin{minipage}{.5\textwidth}
			\centering
			\includegraphics[width=80mm]{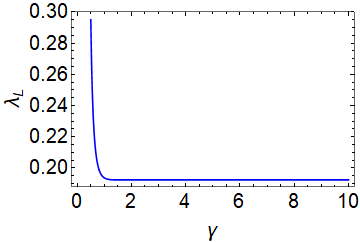}
		\end{minipage}
\caption*{Fig. 7. The angular velocity $\Omega_c $ and the Lyapunov exponent $\lambda_{\rm L}$ with respect to the parameters $\gamma$. Here we set $m=1$ and $q=0.02$.}
\label{figure5}
\end{figure}

\section{Shadows of the new family of ABG black holes}

We investigate the shadow radius of the new family of ABG black holes. The shadow radius of a static spherically symmetric black hole observed by a static observer at infinity takes~\cite{P38,P51,P59,P60} the form,
\begin{equation}
\label{14}
R_{\rm sh}=\frac{1}{\Omega_{c}}=\frac{r_{\rm c}}{\sqrt{f(r_{\rm c})}}.
\end{equation}

In Fig. 8, we plot the shadow radius $R_{\rm sh}$ with respect to the charge $q$ when $\gamma$ is given for different cases. We can see that the shadow radius decreases monotonically as the charge $q$ increases, which is inverse to the behaviour of the real parts as shown in Fig. 4 and Fig. 6.

\begin{figure}
		\centering
\begin{minipage}{.5\textwidth}
			\centering
			\includegraphics[width=80mm]{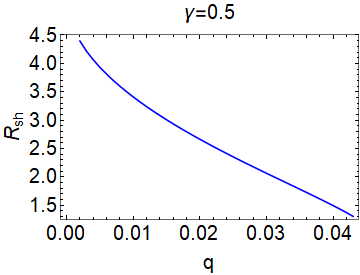}
		\end{minipage}%
		\begin{minipage}{.5\textwidth}
			\centering
			\includegraphics[width=80mm]{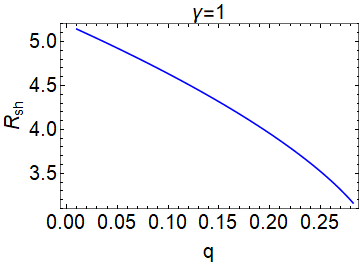}
		\end{minipage}
\begin{minipage}{.5\textwidth}
			\centering
			\includegraphics[width=80mm]{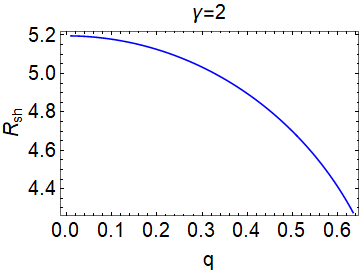}
		\end{minipage}%
		\begin{minipage}{.5\textwidth}
			\centering
			\includegraphics[width=80mm]{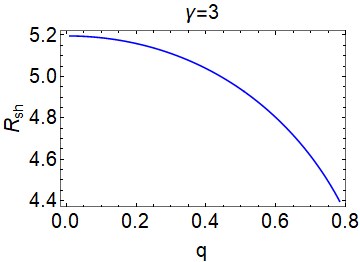}
		\end{minipage}
\begin{minipage}{.5\textwidth}
			\centering
			\includegraphics[width=80mm]{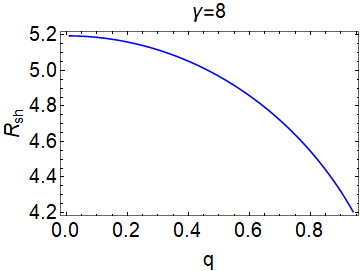}
		\end{minipage}%
		\begin{minipage}{.5\textwidth}
			\centering
			\includegraphics[width=80mm]{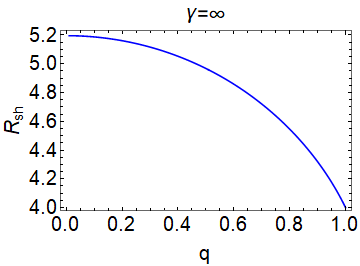}
		\end{minipage}

		\caption*{Fig. 8. The shadow radius $R_{\rm sh}$ with respect to the charge $q$ when $\gamma$ is given for different cases, where the case of $\gamma=2$ corresponding to the ABG black hole and the case of $\gamma\to\infty$ to the Reissner-Nordstr\"om black hole are attached for comparison. Here we set $m=1$.}
\label{figure6}
\end{figure}

In Fig. 9, we plot the shadow radius $R_{\rm sh}$ with respect to the parameter $\gamma$ for a fixed charge,  $q=0.02$. We can see that the shadow radius increases rapidly at first and then almost remains unchanged as the  parameter $\gamma$ increases, which shows that the parameter $\gamma$  has a significant impact on the shadow radius when it is less than one. This is inverse to the behaviour of the real parts as shown in Fig. 5 and Fig. 7.

\begin{figure}
\centering
\begin{minipage}[t]{0.8\linewidth}
\centering
\includegraphics[width=100mm]{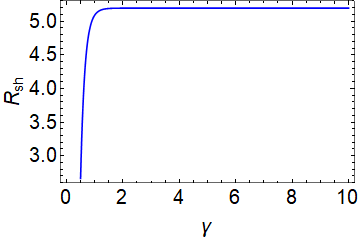}
\caption*{Fig. 9. The shadow radius $R_{\rm sh}$ with respect to the parameter $\gamma$. Here we set $m=1$ and $q=0.02$.}
\label{fig7}
\end{minipage}
\end{figure}

In the remaining of this section, we think that it is necessary to investigate constraints  on  the new family of ABG black holes in terms of the  shadow data of  the  supermassive  black hole  in  the center of the  M87  galaxy  detected by  the  EHT. Based on the observations~\cite{P16,P80,P81,P82}, the angular size of the shadow of the  $M87^{*}$  black  hole is $\delta =(42\pm 3)$ $\mu as$ (microarcsecond), the distance to the  black  hole  $D=16.8_{-0.8}^{+0.8}$ Mpc (million parsec), and the black hole mass $m=(6.5\pm 0.7)\times 10^{9}M_{\bigodot }$. As a result, the shadow diameter $d_{M87^{*}}$  in terms of the unit of mass  is~\cite{P80,P81}
\begin{equation}
d_{M87^{*}}\equiv \frac{D\delta }{m}\approx 11.0\pm 1.5,   %\tag{19}
\end{equation}
which means that  the range of  the shadow diameter is  $9.5 \lesssim d_{M87^{*}}=2R_{sh}\lesssim  12.5$  in the  $1\sigma $ confidence region and  $8.0 \lesssim d_{M87^{*}}=2R_{sh}\lesssim  14.0$  in the  $2\sigma $ confidence region.

We plot the allowed regions  in  the parameter space ($\gamma$, $q$)  for  the new family of ABG black holes in accordance with the  shadow diameter  $ d_{M87^{*}}$  in the  $1\sigma $ and  $2\sigma $ confidence regions  in Figs. 10 and 11, respectively,  in which we also give the  range  of the angular velocity $\Omega_c $ that  connects the real part of  the quasinormal mode frequencies $\omega _{l\gg 1}$  in  the eikonal limit  in the parameter space ($\gamma$, $q$).

\begin{figure}
		\centering
\begin{minipage}{.5\textwidth}
			\centering
			\includegraphics[width=80mm]{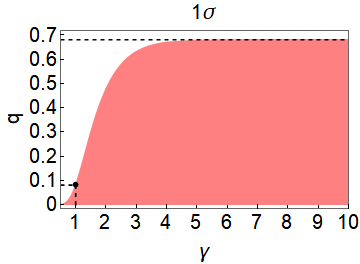}
		\end{minipage}%
		\begin{minipage}{.5\textwidth}
			\centering
			\includegraphics[width=80mm]{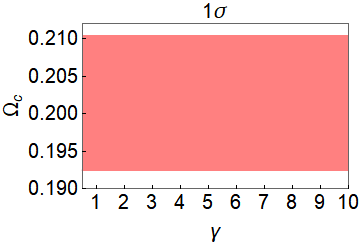}
		\end{minipage}
		\caption*{Fig. 10.   The area shaded in pink  on the left shows  the allowed region  in  the parameter space ($\gamma$, $q$)  for  the new family of ABG black holes, which is plotted in accordance with the shadow diameter of  the  $M87^{*}$  black  hole detected by  the  EHT  in the  $1\sigma $ confidence region. The upper dotted line indicates the allowable maximum charge, 0.6806, for the case of $\gamma\to\infty$, and the lower one gives the corresponding value, 0.0809,  for the case of $\gamma=1$, where the black  point is located at  $(1, 0.0809)$.
 The  shaded  area on the right gives the range of the allowable angular velocity $\Omega_c $, $0.1925<\Omega_c\leqslant 0.2105$, for all values of $\gamma$, where the equality corresponds to the case of the upper
limit of the charge q under different values of $\gamma$ in the left graph. Note that the angular velocity is calculated within the parameter space ($\gamma$, $q$) fixed by the left graph. Here we set $m=1$.}
\label{figure4}
\end{figure}

\begin{figure}
		\centering
\begin{minipage}{.5\textwidth}
			\centering
			\includegraphics[width=80mm]{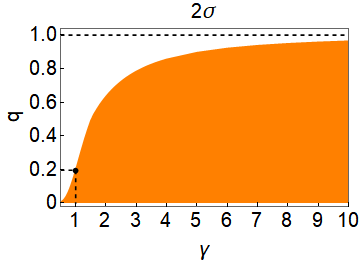}
		\end{minipage}%
		\begin{minipage}{.5\textwidth}
			\centering
			\includegraphics[width=80mm]{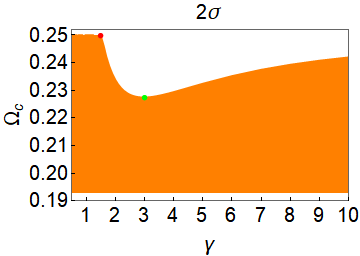}
		\end{minipage}
		\caption*{Fig. 11.  The area shaded in  orange  on the left shows  the allowed region  in  the parameter space ($\gamma$, $q$)  for the new family of ABG black holes, which is plotted in accordance with the shadow diameter of  the  $M87^{*}$  black  hole detected by  the  EHT  in the  $2\sigma $ confidence region.  The upper dotted line indicates the allowable maximum charge, 1, for the case of $\gamma\to\infty$, and the lower one gives the corresponding value, 0.1941,  for the case of $\gamma=1$, where the black  point is located at  $(1, 0.1941)$.
The  shaded  area on the right gives the range of the allowable angular velocity $\Omega_c $, where the parameter space ($\gamma$, $q$) has been fixed in the left graph. The red point is located at  $(1.5, 0.25)$ and the green one at  $(3, 0.2275)$. The range of the allowable angular velocity is $0.1925<\Omega_c\leqslant 0.25$  for  the cases of $0<\gamma\leqslant 1.5$ and $\gamma\to\infty$, where the equality corresponds to the case of the upper limit of the charge q under different values of $\gamma$ in the left graph. Here we set $m=1$.}
\label{figure4}
\end{figure}

In the  $1\sigma $ confidence region,  we can see  from Fig. 10  that  the increase of $\gamma$  makes the allowable range of  $q$ become large. Especially for the new family of ABG black holes  with $0<\gamma<1$, the allowed charge $q$ is less than 0.0809, while for this family of  black holes with $1<\gamma<\infty$, the allowed charge $q$ is less than 0.6806.  In addition, we can find that the range of  the  allowable  angular velocity $\Omega_c $ of  the  new  black holes  is $0.1925<\Omega_c\leqslant 0.2105$ for all  values of  $\gamma$, where the equality corresponds to the case of the upper
limit of the charge q under different values of $\gamma$ in the left graph. The right graph implies that it is impossible to impose constraints on  the parameter $\gamma$  in terms of  $\Omega_c $ or $R_{sh}$.% which equals the reciprocal of  $\Omega_c $.

In the  $2\sigma $ confidence region,  we can see  from Fig. 11  that  the increase of the parameter $\gamma$   makes the allowable range of  the charge $q$ become large. Especially for the new family of ABG black holes  with $0<\gamma<1$, the allowed charge $q$ is less than 0.1941, while for this family of black holes with $1<\gamma<\infty$, the allowed charge $q$ is less than 1.  In addition, we can see that the range of  the  allowable  angular velocity $\Omega_c $  is $0.1925<\Omega_c\leqslant 0.25$ for the case of $0<\gamma\leqslant1.5$.
For the case of $1.5<\gamma< \infty$, the increase of $\gamma$ makes the allowable range of the angular velocity decrease to the minimum range at first, $0.1925<\Omega_c\leqslant 0.2275$ at $\gamma=3$, and then increase to the maximum range, $0.1925<\Omega_c\leqslant 0.25$ at $\gamma\to\infty$.

%For $1.5<\gamma\leqslant 3$, the increase of the parameter $\gamma$ makes the angular velocity become small, and the allowable range of  the angular velocity is $0.1925<\Omega_c\leqslant 0.2275$. Finally, for $3<\gamma< \infty$, the increase of the parameter $\gamma$ makes the angular velocity become large, and the allowable range of  the angular velocity is $0.2275<\Omega_c\leqslant 0.25$, where the equality corresponds to the limit of  $\gamma\to\infty$.

In  Figs. 10 and 11 the black, red, and green points are associated with special values of $\gamma$. Now we use the sixth-order WKB approximation method  to calculate the fundamental quasinormal mode frequencies of massless scalar  field perturbations  by taking such a range of $\gamma$ that includes these special points. The results are shown in Table 1.
\begin{table}[H]
\centering
\caption{The quasinormal mode frequencies of massless scalar  field perturbations for the new family of ABG black holes  are computed in the allowed regions of the parameter space ($\gamma$, $q$) in the  $1\sigma $ and $2\sigma $ confidence regions.} 
\label{table11}
\begin{tabular}{|c|c|c|c|c|c|c|c|c|c|c|c|c|c|}
\hline
  \multicolumn{4}{|c|}{$m=1$, $l=1$, $n=0$}  \\ \hline

 \multirow{2}[0]{*}{$\gamma$}   &  $1\sigma $   or   $2\sigma $& $1\sigma $  &  $2\sigma $                   \\  \cline{2-4}
&   $\omega_{min}$  &   $\omega_{1max}$ &    $\omega_{2max}$	      \\  \cline{1-4}
0.5 &   \multirow{11}[0]{*}{0.29291 $-$ 0.0977616$i$} &    0.320407  $-$ 0.103985$i$ &   0.380271  $-$ 0.116588$i$  \\  \cline{1-1} \cline{3-4}
1 &	  	 &    0.320705  $-$ 0.100095$i$  &  0.380517  $-$ 0.101529$i$    \\  \cline{1-1} \cline{3-4}
1.5 &   	 &   0.321042  $-$ 0.0959725$i$  &  0.377801 $-$ 0.0816016$i$     \\  \cline{1-1} \cline{3-4}
2 &        &  0.321066 $-$ 0.0940529$i$   &   0.353446 $-$ 0.0814751$i$   	\\  \cline{1-1} \cline{3-4}
3 &	     &  0.320781 $-$ 0.095721$i$   &   0.343584 $-$ 0.0862216$i$  	 \\  \cline{1-1} \cline{3-4} 	
4 &	     &  0.320749 $-$ 0.0978395$i$  &   0.346517 $-$ 0.0895958$i$ 	 \\  \cline{1-1} \cline{3-4}
5 &	     &  0.320819 $-$ 0.0987367$i$   &   0.351123 $-$ 0.0912991$i$ 	 \\  \cline{1-1} \cline{3-4}
6 &	     &  0.320836 $-$ 0.0990882$i$   &   0.356101 $-$ 0.0918408$i$	 \\  \cline{1-1} \cline{3-4}
7 &	    &   0.320808 $-$ 0.0992539$i$    &   0.360232 $-$ 0.0921007$i$ 	 \\  \cline{1-1} \cline{3-4}
8 &	    &   0.320776 $-$ 0.0993385$i$    &   0.361658 $-$ 0.0929356$i$ 	 \\  \cline{1-1} \cline{3-4}
$\infty $ & &  0.320765 $-$ 0.0993989$i$ &    0.377706 $-$ 0.0893423$i$ 	 \\  \cline{1-4}
\end{tabular}
\end{table}

Moreover, by taking $m=6.5\times 10^{9}M_{\bigodot }$ and using the formula,
\begin{equation}
f=\frac{{\rm Re}(\omega) }{2\pi m}\times \frac{c^{3}}{G},  % \tag{20}
\end{equation}
we convert the oscillation  frequencies  ${\rm Re}(\omega)$ in Table 1 to their values $f$  in terms of the unit of  Hertz (Hz). The results are shown in Table 2.
\begin{table}[H]
\centering
\caption{ The oscillation  frequencies  $f$  of massless scalar  field perturbations for the new family of ABG black holes  are computed in the  allowed regions  of the parameter space ($\gamma$, $q$) in the  $1\sigma $ and $2\sigma $ confidence regions.} 
\label{table12}
\begin{tabular}{|c|c|c|c|c|c|c|c|c|c|c|c|c|c|}
\hline
  \multicolumn{4}{|c|}{$m=6.5\times 10^{9}M_{\bigodot }$, $l=1$, $n=0$}  \\ \hline

 \multirow{2}[0]{*}{$\gamma$}   &  $1\sigma $  or   $2\sigma $& $1\sigma $  &  $2\sigma $                   \\  \cline{2-4}
&   $f_{min}$ $(10^{-6}Hz)$ &   $f_{1max}$ $(10^{-6}Hz)$  &    $f_{2max}$	$(10^{-6}Hz)$      \\  \cline{1-4}
0.5 &   \multirow{11}[0]{*}{1.456} &    1.593 &  1.890  \\  \cline{1-1} \cline{3-4}
1 &	  	 &    1.594  &  1.891    \\  \cline{1-1} \cline{3-4}
1.5 &   	 &   1.596  &  1.878     \\  \cline{1-1} \cline{3-4}
2 &        &  1.596   &   1.757  	\\  \cline{1-1} \cline{3-4}
3 &	     &  1.594   &   1.708  	 \\  \cline{1-1} \cline{3-4} 	
4 &	     &   1.594  &   1.722 	 \\  \cline{1-1} \cline{3-4}
5 &	     &  1.595   &   1.745 	 \\  \cline{1-1} \cline{3-4}
6 &	     &  1.595   &   1.770	 \\  \cline{1-1} \cline{3-4}
7 &	    &  1.595    &   1.791	 \\  \cline{1-1} \cline{3-4}
8 &	    & 1.594   &   1.798	 \\  \cline{1-1} \cline{3-4}
$\infty $ & &  1.594  &   1.877 	 \\  \cline{1-4}
\end{tabular}
\end{table}

From the two figures and two tables we have the following conclusions:
\begin{itemize}
\item  The frequency range of the fundamental mode with  $l=1$ matching the shadow diameter of  the  $M87^{*}$  black  hole  detected by EHT is about $1.4\times 10^{-6}Hz\sim 1.9\times 10^{-6}Hz$.
\item In the $1\sigma $  confidence region,  it is extremely difficult to constrain $\gamma$ by using the quasinormal mode frequencies.  While  in the $2\sigma $  confidence region,  it is possible to exclude certain range of  $\gamma$ in terms of the  quasinormal mode frequencies.  For example, if $f$ is about $1.8\times 10^{-6}Hz\sim 1.9\times 10^{-6}Hz$, then $\gamma$ $\notin$ [2, 8].
\end{itemize}

\section{Conclusion }

In this paper, we construct a type of ABG related black holes with five parameters and confirm its regularity if the conditions: $\alpha \gamma \geqslant 6$, $\beta \gamma \geqslant 8$, and $\gamma >0$, are satisfied. Then we just consider the saturated case of these conditions, i.e. $\alpha=6/\gamma$ and $\beta= 8/\gamma$, and give a new family of ABG black holes associated with only three parameters, $m$, $q$, and $\gamma$. For this new family of ABG black holes, we investigate its thermodynamic and dynamic properties, such as Hawking temperature, quasinormal mode frequency, and shadow radius. By comparing the properties with those of the ABG black hole and Reissner-Nordstr\"om black hole, we determine the characteristics of the new ABG model. Finally,  we use the shadow data of  the $M87^{*}$ black  hole detected by  the EHT  to give some constraints  on  the new family of ABG black holes. Here we summarize the major ones.
\begin{itemize}
\item The behaviour of the Hawking temperature with respect to the horizon radius is similar to that of the ABG black hole and Reissner-Nordstr\"om black hole. For a fixed $q$, the maximum Hawking temperature depends on the parameter $\gamma$, more precisely, it is less than that of the ABG black hole when $\gamma$ is less than 2, and larger than that of the ABG black hole when $\gamma$ is larger than 2. In any case, the maximum Hawking temperature is less than that of the Reissner-Nordstr\"om black hole, but their difference becomes small when $\gamma$ is large.

\item The behaviour of the real parts of quasinormal modes with respect to charge $q$ is similar to that of the ABG black hole and Reissner-Nordstr\"om black hole. This shows that the oscillation law of the new ABG black hole is similar to that of the ABG black hole and Reissner-Nordstr\"om black hole.

\item The behaviour of the negative imaginary parts of quasinormal modes with respect to charge $q$ is completely different from that of the ABG black hole but similar to that of the Reissner-Nordstr\"om black hole. This means that the decay law of the new ABG black hole is completely different from that of the ABG black hole but
    similar to that of the Reissner-Nordstr\"om black hole.
\item The behaviour of the shadow radii with respect to charge $q$ is similar to that of the ABG black hole and Reissner-Nordstr\"om black hole. This is obvious because the shadow radius is just the reciprocal of the real parts of quasinormal modes under certain conditions~\cite{P38,P51,P59,P60}.

\item The shadow data  of  the $M87^{*}$ black  hole detected by the EHT impose  an upper limit on the charge $q$  of the new family of ABG black holes.  This upper limit increases rapidly at first and then slowly but does not exceed the mass of the $M87^{*}$ black  hole at last when the parameter $\gamma$ is increasing and going to infinity. In addition, the data restrict the frequency range of the fundamental mode with  $l=1$ to $1.4\times 10^{-6}Hz\sim 1.9\times 10^{-6}Hz$.

\end{itemize}

\section*{Acknowledgments}

This work was supported in part by the National Natural Science Foundation of China under Grant No. 11675081.  The authors would like to thank the anonymous referee for the helpful comments that improve this work greatly.

%\newpage

\end{document}